\documentclass{osa-article}

\journal{osac}

\articletype{Research Article}

\usepackage{lineno}
\usepackage{comment}
\usepackage[normalem]{ulem}
\usepackage{graphicx}
\usepackage{subcaption} 

\begin{document}

\title{Fabrication of a Monolithic 5-Meter Aluminum Reflector for Millimeter-Wavelength Observations of the Cosmic Microwave Background}

\author{Tyler Natoli\authormark{1,2,*}, Bradford Benson\authormark{3,1,2}, John Carlstrom\authormark{1,2,4}, Eric Chauvin\authormark{5}, Bruno Clavel\authormark{6}, Nick Emerson\authormark{7}, Patricio Gallardo\authormark{2}, Mike Niemack\authormark{8,9}, Steve Padin\authormark{10}, Klaus Schwab\authormark{11}, Lutz Stenvers\authormark{12}, and Jeff Zivick\authormark{1}} 

\address{\authormark{1} Department of Astronomy and Astrophysics, University of Chicago, 5640 South Ellis Avenue, Chicago, IL, 60637, USA\\
\authormark{2} Kavli Institute for Cosmological Physics, University of Chicago, 5640 South Ellis Avenue, Chicago, IL, 60637, USA\\
\authormark{3} Fermi National Accelerator Laboratory, MS209, P.O. Box 500, Batavia, IL, 60510, USA\\
\authormark{4} Department of Physics, University of Chicago, 5640 South Ellis Avenue, Chicago, IL, 60637, USA\\ 
\authormark{5} Eric Chauvin Consulting Engineer, Suite 501-117, 980 Birmingham Road, Milton, GA, 30004, USA\\
\authormark{6} CadCao, 3 le petit bois, 268 Route du Mollard, 38560 Jarrie, France\\
\authormark{7} Department of Astronomy/Steward Observatory, University of Arizona, Tucson, AZ 85721, USA\\
\authormark{8} Department of Physics, Cornell University, Ithaca, NY 14853, USA \\
\authormark{9} Department of Astronomy, Cornell University, Ithaca, NY 14853, USA \\
\authormark{10} California Institute of Technology, 1200 East California Boulevard., Pasadena, CA, 91125, USA\\
\authormark{11} CONCAD GmbH, Heidingsfelder Weg 10, 74731 Wallduern, Germany\\
\authormark{12} mtex antenna technology gmbh, Berta-Cramer-Ring 32a, 65205 Wiesbaden, Germany\\
}

\email{\authormark{*}tnatoli@uchicago.edu} 

\begin{abstract}
We have demonstrated the fabrication of a monolithic, 5-meter diameter, aluminum reflector with 17.4 $\mu$m RMS surface error. 
The reflector was designed to avoid the problem of pickup due to scattering from panel gaps in a large, millimeter-wavelength telescope that will be used for measurements on the cosmic microwave background. 
\end{abstract}

\section{Introduction}

Large-aperture telescopes for millimeter-wave measurements of the cosmic microwave background (CMB) have historically used reflectors made of close-packed, ~1 meter-scale panels \cite{carlstrom11,padin08,fowler07}. 
The gaps between these panels scatter light and create sharp features in the side lobes of the telescope beam at large angular scales \cite{gudmundsson21}. 
Using a monolithic reflector\footnote{Here we use 'monolithic reflector' to indicate a reflector that has no gaps after machining but is not necessarily made from a single piece of metal. This is an analog to monolithic glass reflectors that are made of multiple fused cells but formed into a single reflector \cite{yoder02}} will reduce the scattering in the beam by eliminating the gaps between panels.

Until now the surface error of a monolithic aluminum reflector viable at millimeter wavelengths for a large-aperture telescope was unknown and believed to be too large for millimeter-wavelength CMB measurements. 

Here we demonstrate that a monolithic, 5-meter aluminum reflector can be manufactured with a large scale root-means-square (RMS) error of 17.4 $\mu$m.

\section{Reflector Design}

\begin{figure}
    \begin{subfigure}{0.5\textwidth}
        \includegraphics[trim={0 40 0 80},clip,width=\textwidth]{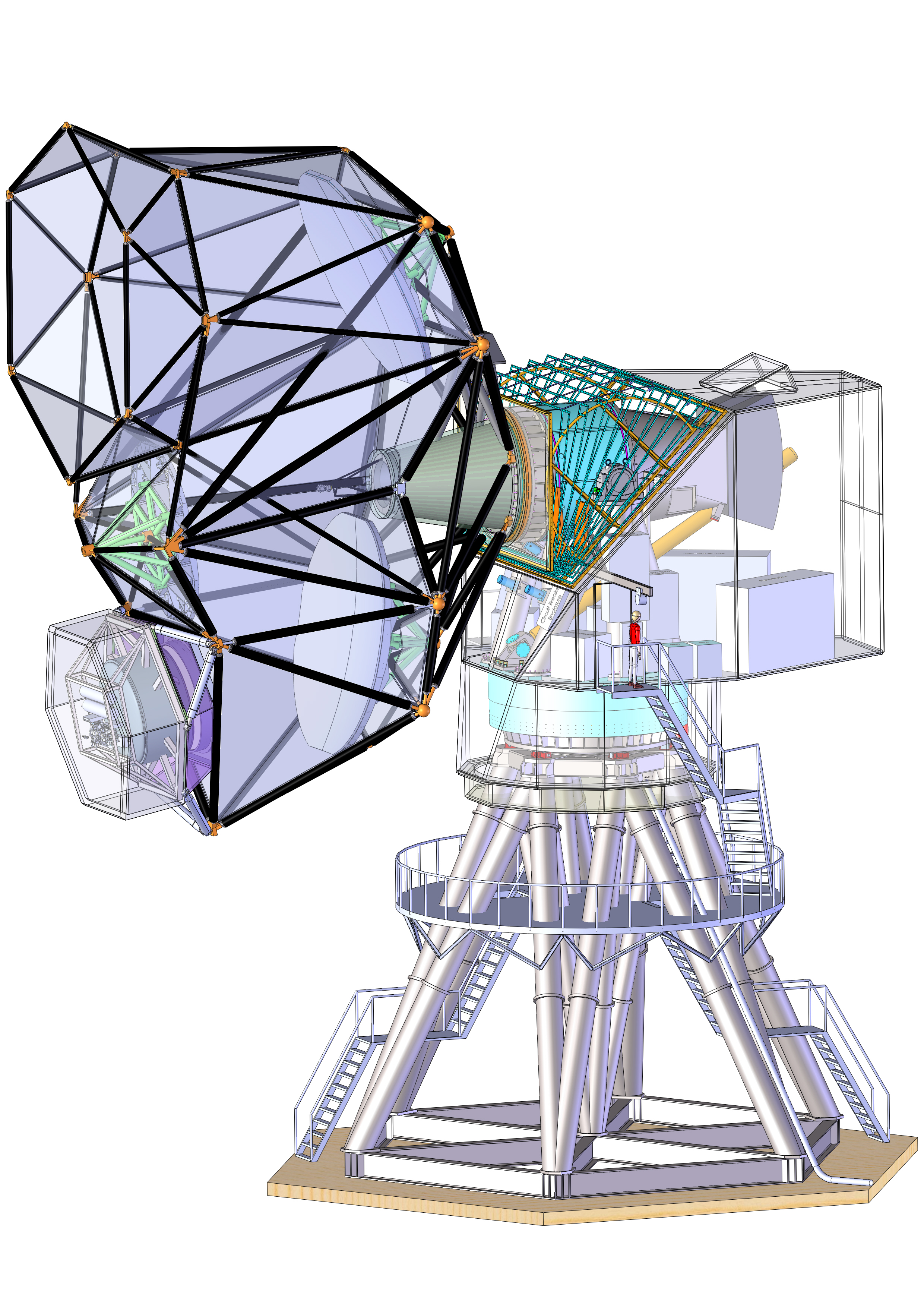}
        \caption{Telescope Design}
        \label{fig:TMA_full}
    \end{subfigure}
    \hfill
    \begin{subfigure}{0.45\textwidth}
        \includegraphics[width=\textwidth]{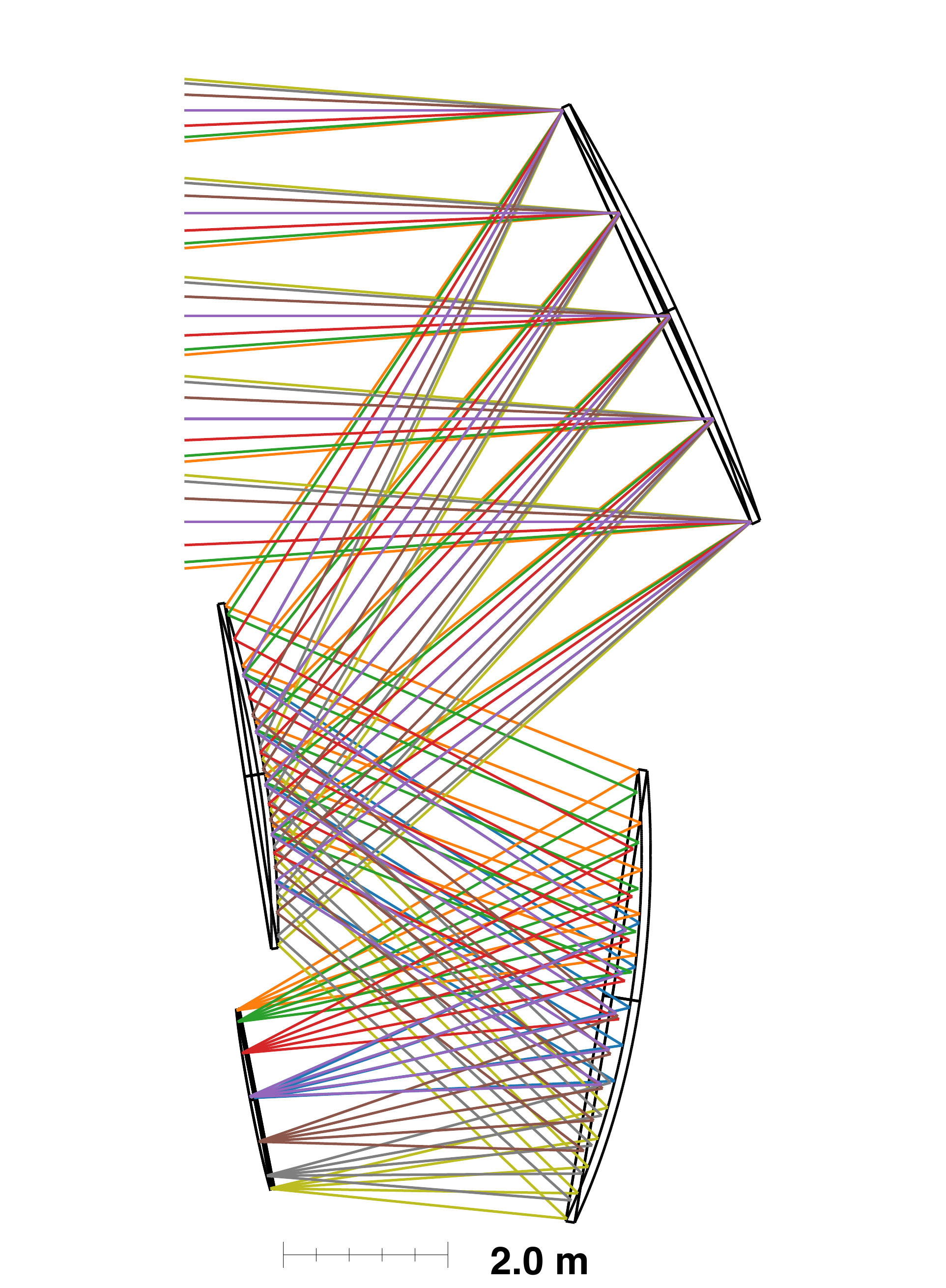}
        \caption{Optics Design}
        \label{fig:ray_trace}
    \end{subfigure}
    \caption{(a) The intended three-mirror anastigmat telescope that will house the reflector discussed in this paper. 
    In the orientation shown the telescope would be observing near the horizon. 
    The reflector fabricated is the primary reflector and is the shown at the top of the model. 
    The carbon fiber space-frame is shown in black and the semi-transparent panels between the carbon fiber struts will be opaque baffling to block stray light from entering the receiver seen on the lower left of the tipping structure. 
    Note the 1.8-meter human figure with a red shirt halfway up the structure for a size reference. 
    (b) The optics design of the three-mirror anastigmat telescope.
    The primary and tertiary reflectors (top and bottom reflectors) are concave and the secondary reflector is convex.}
    \label{fig:STMA_image}
\end{figure}

The 5-meter reflector described in this paper is a prototype primary for the three-mirror anastigmat (TMA) telescope shown in Figure \ref{fig:STMA_image} and discussed in \cite{padin18,gallardo22}.
We will outline the telescope design briefly here, but refer to \cite{padin18} for more details concerning telescope design and a detailed discussion of the benefits of using monolithic reflectors in CMB telescopes. 

TMA telescope designs are able to achieve generous fields of view by correcting for all the major Seidel aberrations. 
The mapping speeds of current CMB experiments are primarily set by the number of detectors with unique sky positions making simultaneous observations. 
The field of view for the TMA telescope design as currently planned is 9 degrees in diameter at a wavelength of 1.1 mm \cite{gallardo22}.

The surface equation for the 5-meter monolithic reflector discussed in this paper is
\begin{multline}\label{eq:M1_surface}
z(x,y) = C_{X0Y1}\frac{y}{R}+C_{X2Y0}(\frac{x}{R})^2+C_{X0Y2}(\frac{y}{R})^2+C_{X2Y1}(\frac{x}{R})^2 (\frac{y}{R}) \\
+C_{X0Y3}(\frac{y}{R})^3+C_{X4Y0}(\frac{x}{R})^4+C_{X2Y2}(\frac{x}{R})^2 (\frac{y}{R})^2+C_{X0Y4}(\frac{y}{R})^4.
\end{multline} 
Where $R$ is a normalization factor and each of the eight $C_{X*Y*}$ are constants describing the desired prescription. 
This surface equation is an even function about the symmetry plane of the TMA, which is why it does not have every powered term of $x$ and $y$, and has sufficient terms to describe a 5-meter reflector that operates well at a wavelength of 1 mm.
For information about how the prescription was chosen and a more in depth discussion of the optics design see [Gallardo et al. in prep].

\begin{figure}
\centering\includegraphics[width=6cm]{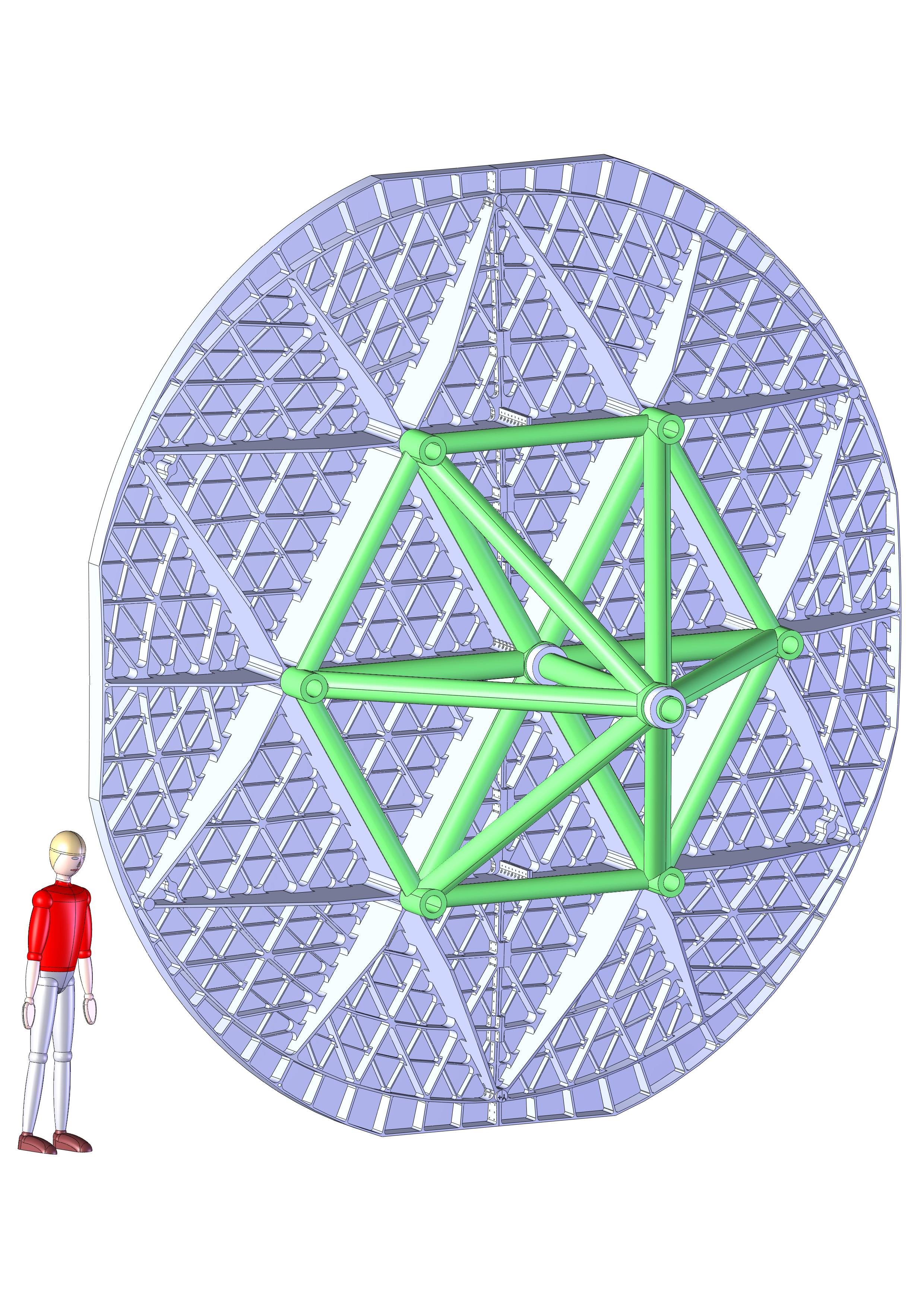}
\caption{A model of the back of the fabricated reflector with its spaceframe support cone shown in green. The human figure for scale is 1.8-meters tall. 
}
\label{M1_with_cone}
\end{figure}

When installed in the telescope the reflector will be supported by a spaceframe cone at six equally-spaced radially-compliant support points on the back of the reflector as seen in Figure \ref{M1_with_cone}.
This spaceframe cone will provide a stiff support for the reflector and will be made of aluminum (like the reflector) so that it will have the same coefficient of thermal expansion as the reflector.  
While being machined the reflector was bolted to the machine bed at the six cone support points.
The same interface loads experienced at each of the six support points during the final surface measurements will be reproduced when the reflector is installed in the telescope. 

The overall reflector shape is approximately oval with minor and major axis diameters of 5 meters and 5.54 meters respectively. 
The reflector diameter of $\sim$5 meters was chosen to maintain a Strehl ratio $\geq 80 \%$ at 1 mm wavelengths given reasonable assumptions about thermal deformations \cite{padin18}. 
The reflector is only half a meter thick with a light-weighted back and a surface skin thickness of 8 millimeters to yield small gravitational deformations.

Small scale scattering off a reflector's surface can lead to excess detector loading as the wide angle scattered light can be incident on hotter components than the effective sky temperature. 
To keep the small scale scattering across all three reflectors of the TMA result in less than a 1$\%$ Strehl degradation, each individual mirror must have a small scale surface error of less than 4 $\mu$m. 
Large scale scattering off the surface of a reflector will distort the beam pattern but should not lead to excess detector loading so there is a less stringent requirement on large scale surface errors. 
For the TMA telescope to be diffraction limited at a wavelength of 1 mm and to keep the Strehl ratio above 80$\%$ for the entire TMA, the large-scale surface error of any individual reflector needs to be below 21 $\mu$m. 

\section{CONCAD Manufacturing Facility}\label{CONCAD_CNC_CMM}

The reflector fabrication was engineered and managed by mtex antenna technology\footnote{https://www.mtex-at.com/} based in Wiesbaden, Germany. 
The production of the reflector was performed at CONCAD\footnote{https://www.concad-gmbh.de}, a fabrication shop located in Wallduern, Germany.

Thermal gradients through the reflector need to be kept to less than a few tenths of a Kelvin during cutting and metrology to keep thermal deformations in the reflector under the target surface accuracy. 
This was ensured by the ambient temperature within the CONCAD machining hall being stable to +/- 0.4 $^{\circ}$C over the 3 days all but one of the surface measurements occurred.

The 5-axis milling machine used to cut the reflector from the raw blocks of aluminum was a Starrag FOGS HD 50 150 R75 C with computer numerical control (CNC) from a SIEMENS Sinumerik 840D sl. 
This mill has an active cutting area that is 5 meters wide, 15 meters long, and 2 meters high. 
This milling machine can be used as a coordinate measuring machine (CMM) by replacing the cutting bit on the machine head with a Hexagon 20.50-G-HPP touch probe.

To improve the accuracy of the milling machine/CMM a careful calibration of the machine was independently performed by AfM Technology GmbH\footnote{https://www.afm-tec.info/} in Aalen, Germany per ISO 10360-2:2009 \cite{ISO09}.
This calibration procedure used a laser interferometer system and achieved an RMS calibration error of 4 $\mu$m over the milling machine/CMM region occupied by the reflector during machining and surface metrology.

\section{Reflector Fabrication}

The reflector was fabricated out of two blanks of 5083 aluminum alloy, each measuring 2.53 x 5.45 x 0.54 meters and weighing $\sim$19,800 kg. 
5083 aluminum was chosen for its light weight, good tensile strength, and availability in large enough blanks as it is a castable aluminum.
Using cast aluminum blanks also produces less internal stresses in the final reflector than if an extruded aluminum was used. 
The expected variation in the coefficient of thermal expansion (CTE) across the aluminum blanks will be on the order of 1\% according to measurements in \cite{touloukian70}.
The variation in CTE is a concern as it would lead to peak-to-peak deformations of 10 $\mu$m over the 100 K temperature changes experienced annually at the South Pole, but these deformations are not large enough to spoil the performance of the reflector.

\begin{figure}
\centering\includegraphics[width=6cm]{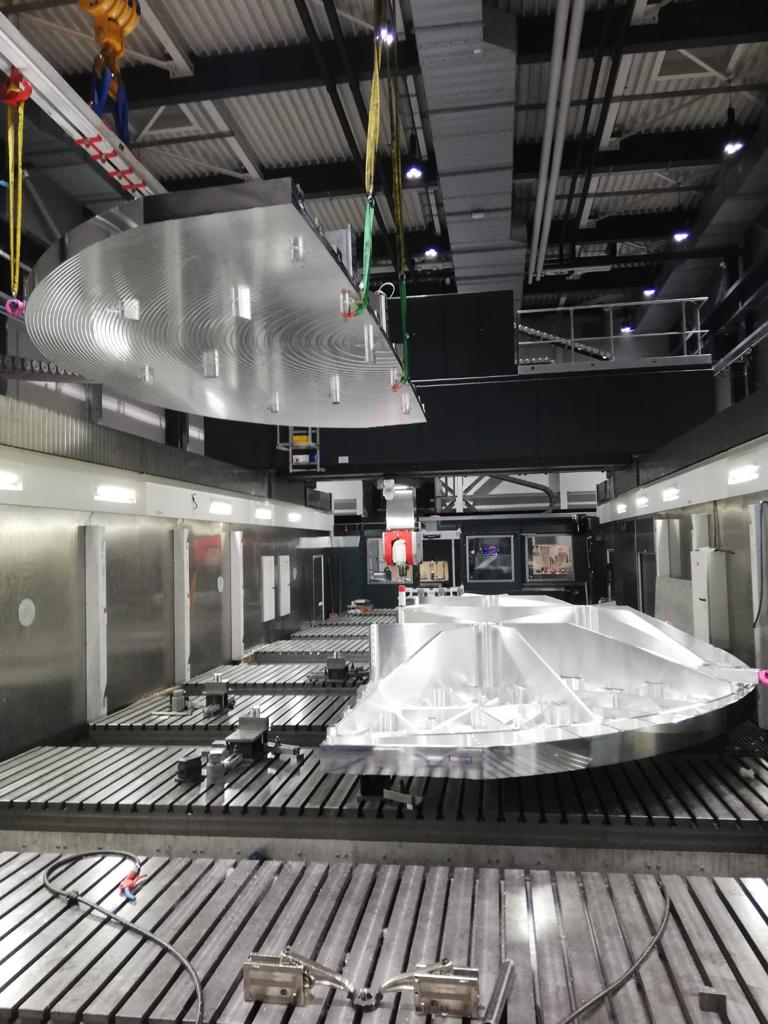}
\caption{The two separate reflector halves after rough machining and before they were bolted together.
The surface of the reflector is pointed downwards and you can see large aluminum pillars on the surface of the reflector that still need to be removed.
}
\label{mirror_halves}
\end{figure}

The aluminum blanks were each rough machined separately into one-half of the final reflector as shown in Figure \ref{mirror_halves}. 
After rough machining the two reflector halves were bolted together along the center line of the reflector using 200 aluminum M8 bolts and 16 slotted keys for alignment. 
After being bolted together the two reflector halves will not be separated and the reflector is treated as one solid object.

The final machining passes over the reflector's surface were done with the reflector halves bolted together and took 20 hours.
Each of the passes during the final surface cutting removed less than 200 $\mu$m of aluminum and the reflector was not polished after machining.
The small-scale surface roughness of the final reflector was measured with a rugosimeter to be less than 1 $\mu$m (Ra).

\begin{figure}
\centering\includegraphics[width=12cm]{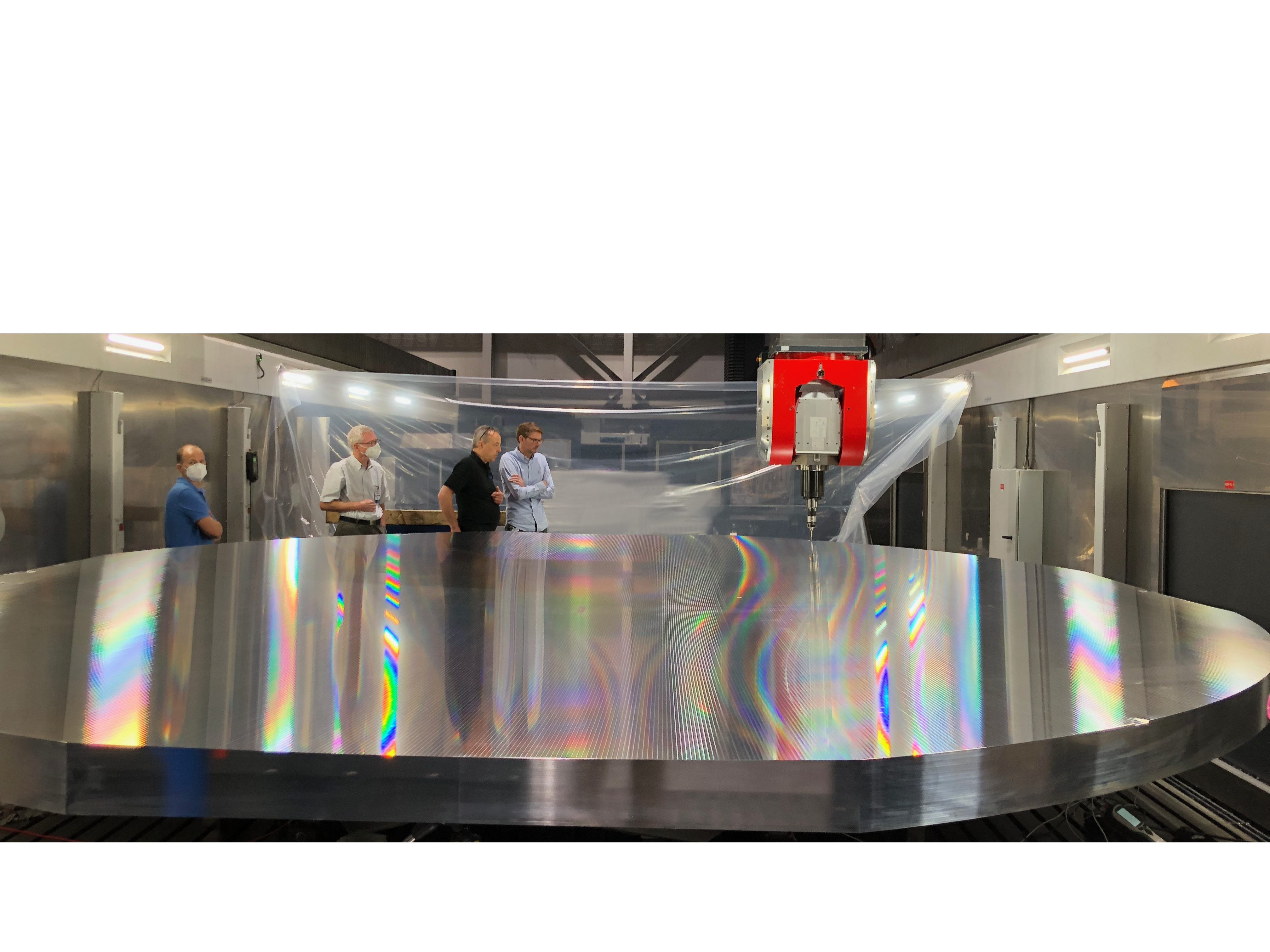}
\caption{The reflector after final machining. 
The milling machine has been transformed into a CMM by replacing the cutting bit with a touch probe to take metrology measurements over the surface of the reflector.}
\label{fig:M1_full_size}
\end{figure}

After machining was finished the reflector's total weight was $\sim$2,122 kg indicating $\sim$95$\%$ of the material from the original aluminum blanks was removed to form the reflector. 
The final machined surface of the reflector is shown in Figure \ref{fig:M1_full_size}. 

\section{Reflector Surface Metrology}

After the reflector fabrication was completed the same milling machine responsible for cutting the reflector was adapted into a CMM as described in Section \ref{CONCAD_CNC_CMM}.
The CMM was used to take measurements over the full surface of the reflector in a grid pattern. 
This grid pattern varied in spacing from 5 cm to 30 cm with a full grid of measurements over the reflector surface taking between 12 hours and 1.5 hours, respectively. 
Refer to Table \ref{metro_measurements_table} for a summary of the reflector configurations and grid pitches for a selection of the surface measurements taken.

\begin{table}[!ht]
    \centering
    \resizebox{1.0\textwidth}{!}{
    \begin{tabular}{|c|c|c|l|}
    \hline
        Measurement & Grid pitch [mm] & Rotation [deg] & Notes \\ \hline
        A & 50 & 0 & Reflector bolted to machine bench. \\ \hline
        B & 300 & 0 & Removed bolts, reflector now on loadcells.\\ \hline
        C & 300 & 0 & Repeated B, this is the "Control" measurement.\\ \hline
        D & 300 & 60 & Reflector rotated 60 deg CCW from control orientation. \\ \hline
        E & 300 & 180 & Reflector rotated 180deg from control orientation. \\ \hline
    \end{tabular}
    }
    \caption{A selection of the reflector surface measurements taken at CONCAD are listed in chronological order. The rotation angle listed is relative to the orientation of the reflector while the final surface was cut.}
    \label{metro_measurements_table}
\end{table}

The initial surface measurement, "A", was taken with the reflector in the same configuration as when it was cut, with the reflector's six support points bolted to the machine bed. 
After this measurement the support point bolts were removed.
All subsequent measurements (B-E in Table \ref{metro_measurements_table}), were made with loadcells at each of the six support points. 
Each loadcell was individually adjusted to provide the same relative heights of the six support points as during final machining. 

In the rest of this section we describe how the surface metrology measurements taken allow us to assign errors to specific components of the fabrication and metrology process; manufacturing error, measurement repeatability, and systematic error.

\begin{figure}
\centering\includegraphics[width=10cm]{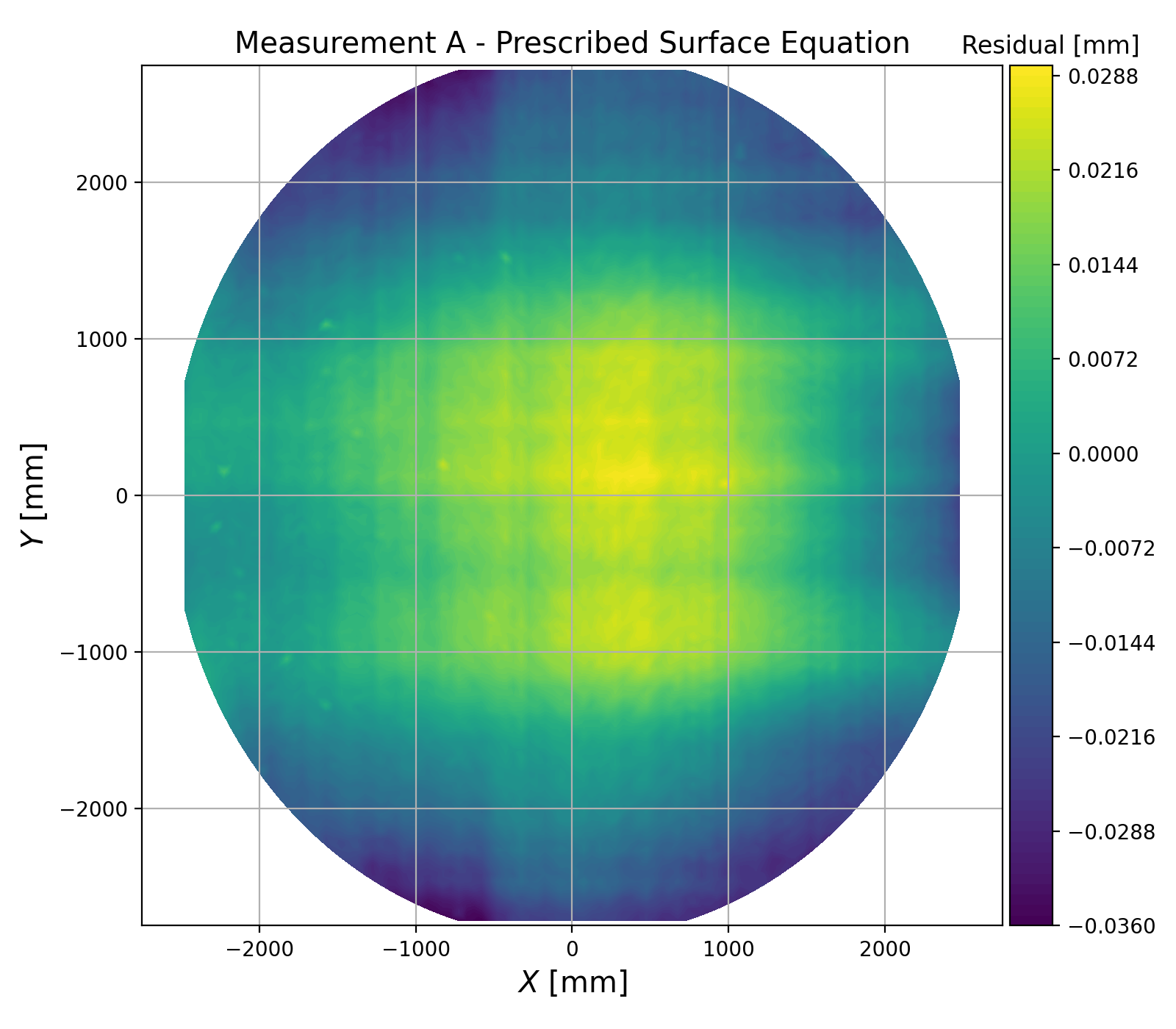}
\caption{The difference between the measured reflector surface (Measurement A in Table \ref{metro_measurements_table}) and the prescribed surface equation. The best-fit plane has been subtracted prior to calculating the RMS and plotting to remove sensitivity to the tip and tilt changing between the reflector fabrication and measurement. The RMS of the difference shown is 14.1 $\mu$m.}
\label{Machining_error_figure}
\end{figure}

\subsection{Manufacturing Error}

There will always be real physical differences between the prescribed surface intended to be produced and the actual reflector surface cut into the aluminum. 
We refer to this difference as the manufacturing error. 

To measure the manufacturing error we performed a surface measurement with the reflector setup unchanged from when the final surface of the reflector was cut, this is measurement A in Table \ref{metro_measurements_table}.
For this measurement the reflector was still on blocks and bolted down to the bed of the milling machine at the six support points. 
Figure \ref{Machining_error_figure} shows the difference between measurement A and the prescribed surface profile. 
After removing the best-fit plane from the difference the RMS is 14.1 $\mu$m.

\subsection{Repeatability}

When experiencing stable environmental conditions, like in the CONCAD machining hall, taking repeated measurements without changing the physical conditions of the measurement allows us to probe the measurement repeatability.

\begin{figure}
\centering\includegraphics[width=10cm]{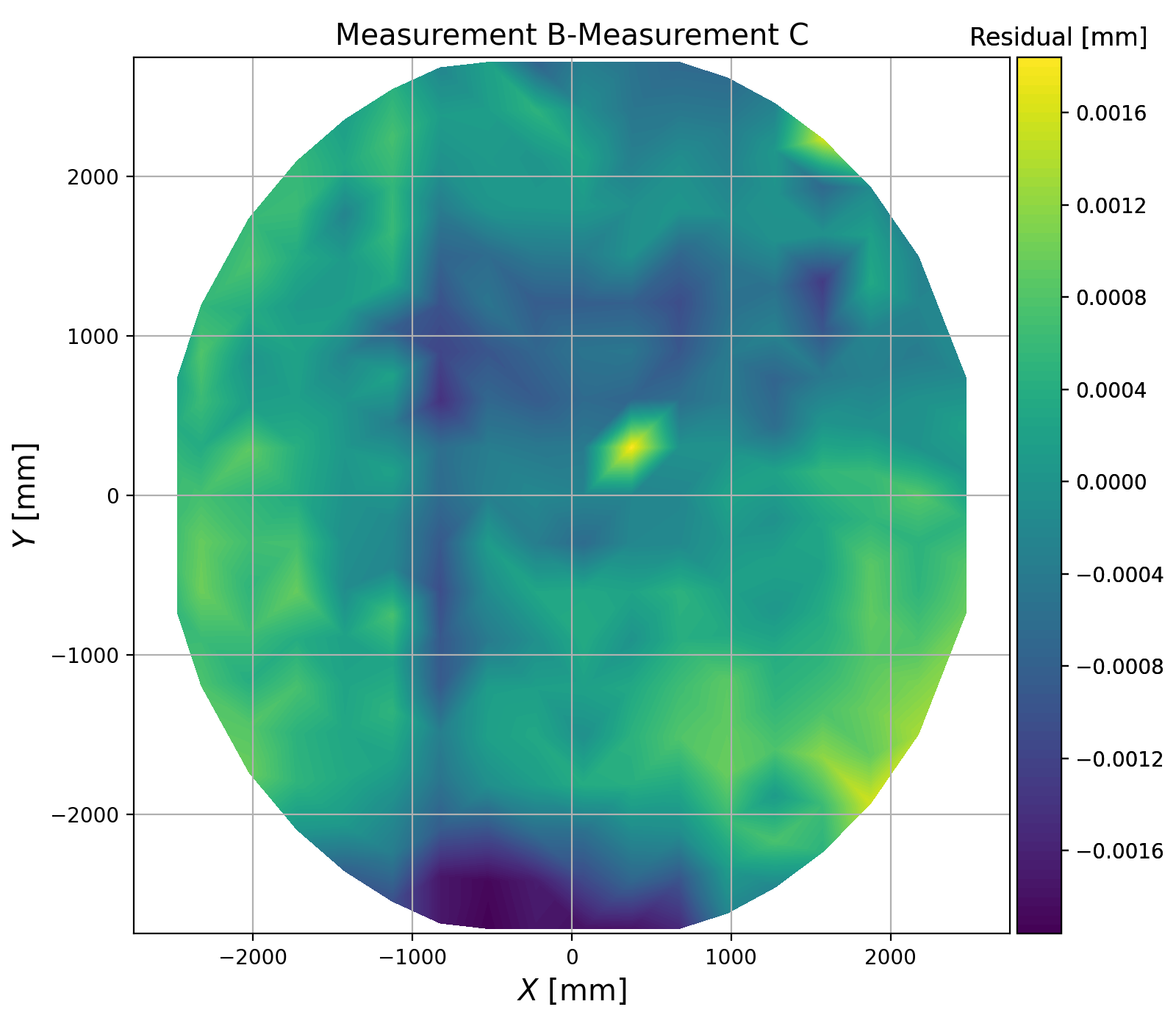}
\caption{The difference between surface measurements B and C from Table \ref{metro_measurements_table} are displayed to show measurement repeatability. 
No changes were made to the measurement or reflector setup between these measurements. 
The RMS of the difference between measurements B and C is 0.7 $\mu$m. 
The best-fit plane has been removed from the difference prior to plotting.}
\label{Repeatability_figure}
\end{figure}

There were four pairs of full surface measurements repeated without making deliberate changes to the physical setup or reflector orientation between each measurement in a pair. 
Figure \ref{Repeatability_figure} shows an example of the difference between one pair of repeated measurements, in this case the two measurements were taken two hours apart. 

The RMS of the differences between each of the four pairs of repeated measurements are all 1.7 $\mu$m or lower.

\subsection{Systematic Error}

As described in Section \ref{CONCAD_CNC_CMM}, the same piece of machinery used to cut the reflector surface was used to perform the surface metrology, the only difference being the machine head insert used. 
With this setup any systematic error in the machine appears in both surface cutting and measuring movements. 
If the reflector is measured in the exact position it was cut in the measurement is blind to any systematic machine errors. 

\begin{figure}
\centering
\resizebox{1.0\textwidth}{!}{
\includegraphics[width=14cm]{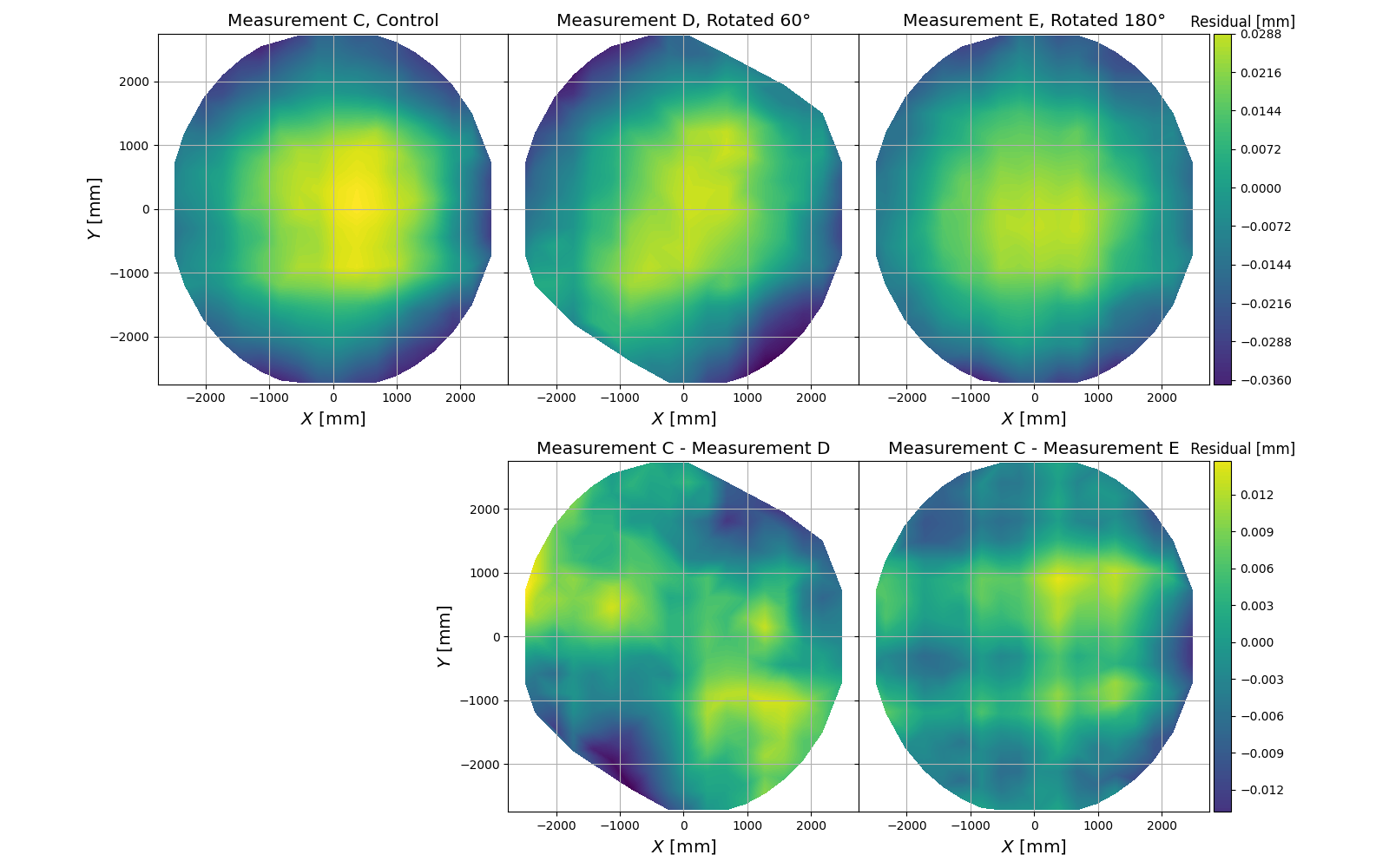}
}
\caption{The top row of plots show surface measurements with the prescribed surface equation and the best-fit plane subtracted off.
The top row of plots all share the colorbar on right of the top row. 
The bottom row shows the difference between the control measurement and the measurements done in rotated orientations with a best-fit plane removed.
The orientation of every plot shown has been rotated to match the control measurement orientation. 
Points measured during measurement D (in the center column) do not span the entire reflector and were limited to areas that could be accessed with the reflector rotated 60 degrees. 
}
\label{Systematic_error_figure}
\end{figure}

To ensure the milling machine and CMM heads trace out different paths we made surface measurements with the reflector in three different rotation orientations; the original Control orientation (same as the cutting orientation), rotated 60 degrees from the Control orientation, and rotated 180 degrees from the Control orientation.
The top row of Figure \ref{Systematic_error_figure} shows these three measurements rotated into the same orientation as the Control measurement. 
We rotate the surface measurements into the same frame prior to calculating the differences to subtract out the physical reflector surface and only probe the measurement systematics. 
In the bottom plots of Figure \ref{Systematic_error_figure} the differences between the the control and each of the other rotations are shown.
Any common features seen in the difference plots indicate systemtatic errors in the machine. 
The average RMS of the two difference plots shown is 6.0 $\mu$m. 

\subsection{Total Surface Error}

The three errors discussed in this section (manufacture, repeatability, and systematic) and the calibration error from Section \ref{CONCAD_CNC_CMM} all need to be combined into a total surface error. 
All of these error terms could be combined in quadrature, which would result in a total RMS error of just under 16 $\mu$m. 
If we had performed surface measurements with the reflector in many different locations and orientations on the milling machine bed we could have produced a systematic error that also captured the calibration error. 
Since we performed a limited number of surface measurements it is not clear how to combine the calibration error and the measured systematic error. 
To be conservative, we add the calibration and systematic errors linearly before combining that sum in quadrature with the manufacturing error and measurement repeatability.
This yields a total large-scale RMS error of 17.4 $\mu$m across the surface of the fabricated 5-meter aluminum reflector. 

\begin{table}[!ht]
    \centering
    \begin{tabular}{|c|c|}
    \hline
        \textbf{Type} & \textbf{RMS [$\boldsymbol\mu$m]} \\ \hline
        Manufacturing error & 14.1 \\ \hline
        Repeatability & 1.7 \\ \hline
        Systematic error & 4.0 \\ \hline
        Calibration error & 6.0 \\ \hline
        Total error & \textbf{17.4} \\ \hline
    \end{tabular}
    \label{error_table}
\end{table}

\section{Conclusion}

We have demonstrated the fabrication of a 5-meter, monolithic, aluminum reflector with a surface error of 17.4 $\mu$m RMS. 
With this surface error the reflector is viable for taking CMB observations at millimeter wavelengths on a large-aperture CMB telescope. 
The ability to fabricate quality monolithic 5-meter reflectors will allow future telescopes to make CMB measurements at larger spatial scales than existing large-aperture CMB telescopes. 

\begin{backmatter}
\bmsection{Disclosures}

The authors declare no conflicts of interest.

\bmsection{Funding}
 
National Science Foundation award number 2034402. 
 
\bmsection{Acknowledgment}
The authors would like to thank the late Richard Hills for his invaluable contributions to the optical prescription of TMA reflector discussed in this work. 

\end{backmatter}

\bibliography{ProtoM1_bib.bib}
 
\end{document}